\documentclass[amsmath, amsfonts, superscriptaddress, twocolumn, prl]{revtex4-1}
\usepackage{graphicx}
\usepackage{epsfig}
\usepackage{bm}
\usepackage{dcolumn}
\usepackage{amsmath}
\usepackage{amssymb}
\usepackage{xcolor}
\usepackage{stackrel}
\usepackage{accents}
\usepackage{latexsym}
\usepackage{verbatim}
\usepackage{hyperref}

\hypersetup{
    colorlinks,
    citecolor=blue,
    filecolor=blue,
    linkcolor=blue,
    urlcolor=blue
}

\newcommand{\beg}{\begin{equation}}
\newcommand{\en}{\end{equation}}

\newcommand{\be}{\begin{eqnarray}}
\newcommand{\ee}{\end{eqnarray}}

\newcommand{\bs}{\begin{equation}\begin{split}}
\newcommand{\es}{\end{split}\end{equation}}

\newcommand{\up}{\uparrow}
\newcommand{\dn}{\downarrow}

\newcommand{\eref}[1]{Eq.~(\ref{#1})}

\begin{document}

\title{Comment on ``Nonequilibrium dynamics of superconductivity in the attractive Hubbard model"}

\author{Maxim Dzero}
\affiliation{Department of Physics, Kent State University, Kent, Ohio 44242, USA}

\author{Emil A. Yuzbashyan}
\affiliation{Department of Physics and Astronomy, Rutgers University, Piscataway, NJ 08854, USA}

\author{Boris L. Altshuler}
\affiliation{Department of Physics, Columbia University, New York, NY 10027, USA}


\maketitle

 In a recent preprint \cite{chern} Chern and Barros report  numerical simulations of the mean-field  interaction quench  dynamics, $U_i\to U_f$, of the attractive  Hubbard model that confirm our earlier prediction~\cite{turbo} of spontaneous eruption of spatial inhomogeneities in the  post-quench state with periodically oscillating superconducting order. However, 
 their interpretation of their own numerics is inconsistent and incomplete, which we point out below.
 
 For spatially uniform states, such as that at the initial stage of the dynamics, the mean-field Hubbard model in Eq.~(2) of Ref.~\onlinecite{chern} is   the  reduced BCS model with 2D tight-binding single-particle  spectrum and renormalized chemical potential, see, e.g., Ref.~\onlinecite{hfb}. Its interaction quench dynamics has been analyzed   in Refs.~\onlinecite{barankov2,tsyplyatev,barankov1,dzero, review}.  If $U_f$ is sufficiently close to $U_i$, the amplitude of the order parameter  asymptotes to a nonzero constant (phase II in   Ref.~\onlinecite{review}).   When $|U_f|$ exceeds a certain $U_i$-dependent threshold, the system goes into phase III, where the order parameter amplitude oscillates periodically and Cooper pairs distribute themselves among two orthogonal Floquet states. In particular,   when $|U|$ is much smaller than the bandwidth $W$, the transition  to phase III occurs at
\beg
\frac{1}{|U_f|}=\frac{1}{|U_i|}-\frac{\pi \nu_F}{2},
\label{main}
\en
where $\nu_F$ is the density of states at the Fermi level  \cite{barankov1,review}. 

As explained in, e.g., Refs.~\onlinecite{turbo,barankov2},  and apparently overlooked by Chern and Barros, Refs.~\onlinecite{barankov2,tsyplyatev,barankov1,dzero, review} address systems smaller than the superconducting coherence length $\xi$. In a bulk system, phase III is unstable with respect to spatial fluctuations \cite{turbo}. Spatial modulations develop through parametric excitations of pairing modes with opposite momenta. Subsequent scattering limits the initial exponential growth, eventually resulting in a random superposition of wave packets of the order parameter of typical size of the order of $\xi$. This effect, termed `Cooper pair turbulence' in Ref.~\onlinecite{turbo}, is similar to the wave turbulence phenomenon in other nonlinear media \cite{sw1,sw2,sw3,sw4}. 
 
 Chern and Barros    attribute the instability of phase III with respect to spatial fluctuations to the large value of $|U_f|$, such that   $\xi$ is of the order of few lattice spacings.  This contradicts \eref{main}, which shows that the system goes into the unstable phase III  for arbitrarily small $|U_f|$  provided $|U_i|$ is sufficiently small. Moreover, in 3D for very large $|U_f|$ and arbitrary $U_i$, the condensate  ends up in the stable phase II \cite{review}, which further undermines this interpretation. The only argument in Ref.~\onlinecite{chern} explaining the instability ``in the large $U_f$ regime" is that each $\Delta_i=\langle c_{i\dn} c_{i\up}\rangle$ ``oscillates with its own amplitude and frequency" leading to the ``Landau-damping of collective superconducting order". This seems especially puzzling, because in the large $|U|$ (atomic) limit of the Hubbard model, all $\Delta_i$, in fact, oscillate with the same frequency $U-2\mu$, as evident from, e.g., Eq.~(4) in Ref.~\onlinecite{chern}. 
 
  Chern and Barros further observe the formation of domain walls in the spatially resolved superconducting order and  find the absence of ``topological defects, or vortices" surprising, while stating earlier that their ``numerical results are consistent with the Cooper pair turbulence phenomenon"  of Ref.~\onlinecite{turbo}.  Yet these features -- domain structure and the absence of vortices -- are typical in spatially nonuniform states arising from a parametric instability \cite{turbo,sw1,sw2,sw3,sw4}. Nevertheless, further work is necessary  to numerically confirm   our predictions of the parametric mechanism of the instability   and 
    for the post-threshold state. In particular, we suggest looking at the momentum distribution of Cooper pairs. This should display an additional peak at the wave-vector of the instability of the order of of $\xi^{-1}$ as the instability starts to develop. At later times, we expect this peak to shift towards larger wave-vectors, signaling  energy cascade to smaller length scales \cite{turbo}. 
  
 It is also important  to be aware  of finite size effects in numerical simulations of interaction quench dynamics of BCS superconductors. These  typically manifest themselves at the timescale  $t_\mathrm{fs}\approx\delta^{-1}$, where $\delta$ is the mean single-particle level spacing \cite{review}. For example, Fig.~1(c) of Ref.~\onlinecite{chern} shows beats in $|\Delta(t)|$ that appear to be a finite size effect.  The authors do not specify the units in which $t$ is measured. Assuming the units are such that the nearest neighbor hopping amplitude is equal to one, we estimate $t_\mathrm{fs}\approx N/W\approx 300$, where $N=48\times48$ is the number of sites. This time is at about the location of the first pronounced minimum in the amplitude of $|\Delta(t)|$ in Fig.~1(c). The behavior of $|\Delta(t)|$ at times $t> t_\mathrm{fs}$ is not representative of a bulk system. Also of concern is the fact that Fig.~1 of Ref.~\onlinecite{chern} consistently shows values of $\Delta_i$ exceeding 1/2. Indeed, any quantum state can be written as $|\psi\rangle=\alpha | 0\rangle+\beta  | 2\rangle+\gamma | 1\rangle$, where the numbers indicate the occupancy of site $i$. It is then straightforward to show that $\Delta_i=\langle\psi| c_{i\dn} c_{i\up}|\psi\rangle=\alpha^*\beta \le 1/2$.
 
 This work was  supported  by the National Science Foundation Grants DMR-1609829 (E. A. Y.)  and DMR-1506547 (M. D.)

\end{document}